\documentclass[a4paper,aps,pra,superscriptaddress,showpacs,twocolumn]{revtex4}

\usepackage{graphicx}
\usepackage{amssymb}
\usepackage{epstopdf}
\usepackage{amsmath}

\usepackage{bm}

\usepackage{ulem}

\newcommand{\half}{\frac{1}{2}}
\newcommand{\up}{\uparrow}
\newcommand{\down}{\downarrow}
\newcommand{\Nv}{N_{\rm V}}

\begin{document}

\title{Vortex lattices for ultracold bosonic atoms in a non-Abelian gauge potential}

\author{Stavros Komineas}
\affiliation{Department of Applied Mathematics, University of Crete, Heraklion, Crete, Greece}
\author{Nigel R. Cooper}
\affiliation{Cavendish Laboratory, University of Cambridge, 19 J J Thomson Avenue, Cambridge CB3 0HE, U.K.}

\begin{abstract}
  The use of coherent optical dressing of atomic levels allows the
  coupling of ultracold atoms to effective gauge fields. These can be
  used to generate effective magnetic fields, and have the potential
  to generate non-Abelian gauge fields.  We consider a model of a gas
  of bosonic atoms coupled to a gauge field with $U(2)$ symmetry, and
  with constant effective magnetic field. We include the effects of
  weak contact interactions by applying Gross-Pitaevskii mean-field
  theory. We study the effects of a $U(2)$ non-Abelian gauge field on the vortex
  lattice phase induced by a uniform effective magnetic field,
  generated by an Abelian gauge field or, equivalently, by rotation of
  the gas.  We show that, with increasing non-Abelian gauge field, the
  nature of the groundstate changes dramatically, with structural
  changes of the vortex lattice. We show that the
  effect of the non-Abelian gauge field is equivalent to the introduction of effective
  interactions with non-zero range. We also comment on the
  consequences of the non-Abelian gauge field for strongly correlated fractional quantum Hall
  states.
\end{abstract}

\maketitle

\section{Introduction}

Atomic Bose-Einstein Condensates (BECs) offer the possibility to study
the physics of quantised vortex lines with unprecedented precision and
control \cite{Fetter_2008,cooper08}.  Experiments on rapidly rotating
gases
\cite{Stock_Battelier_Bretin_Hadzibabic_Dalibard_2005,Madison_Chevy_Wohlleben_Dalibard_2000,Abo-Shaeer_Raman_Vogels_Ketterle_2001,Engels_Coddington_Haljan_Schweikhard_Cornell_2003}
have allowed detailed studies of remarkable states such as large
arrays of vortices, forming vortex lattices. Their static features,
dynamics and response to periodic lattice potentials have been
investigated.

As an alternative to rotation, one can use the dressing by coherent
optical fields to create an effective $U(1)$ gauge potential which
simulates the orbital effects of a magnetic field on a charged
particle \cite{dalibard_gerbier_RMP2011}.  Using such optically induced gauge
fields, the formation of quantized vortices in a rubidium condensate
has been demonstrated in pioneering experimental work
\cite{Lin_Compton_Garcia_Porto_Spielman_2009}.  Optically induced
gauge potentials are not limited to Abelian gauge fields, but can be
naturally extended to the non-Abelian case \cite{dalibard_gerbier_RMP2011}.
There exists a variety of proposed ways to generate non-Abelian gauge
potentials, both in the
continuum \cite{ruseckas05,Jacob_Ohberg_Juzeliunas_Santos_2008} and in
lattice-based systems \cite{osterloh05}

In this paper, we study the consequences of a $U(2)$ non-Abelian gauge
field on the groundstate of a weakly interacting atomic BEC. We focus
on a gauge-field configuration in which the effective magnetic field
is constant in space, and for which there exists a simple exact
solution for the single particle
wavefunctions \cite{Burrello_Trombettoni_2010,Burrello_Trombettoni_2011,estienne11,Palmer_Pachos_2011}.  In the case of a
uniform Abelian magnetic field (or for uniform rotation of the gas)
the spectrum has the Landau level structure.  In this case, for weak
repulsive interparticle interactions the bosonic atoms occupy the
lowest energy Landau level and the mean-field ground state is
well-known to be a lattice of vortices with triangular symmetry.  It
is interesting to ask how the addition of a constant non-Abelian
magnetic field affects this groundstate.  

We show that the effects of the additional non-Abelian magnetic field
can be understood in terms of a change in the effective interatomic
interaction potential, and are equivalent to the effects of an
interaction potential with non-zero range.  Specifically, the effects
of the non-Abelian gauge field are completely encoded on the Haldane
pseudopotentials that describe the interatomic interactions in the
lowest energy single-particle states.  We study the consequences on
the condensed vortex lattice phases by applying Gross-Pitaevskii
mean-field theory to the weakly interacting gas. We show that the
nature of the ground state changes dramatically, with structural
changes in the symmetry of the vortex lattice brought about with
increasing non-Abelian gauge field. We show that these changes are precisely analogous to the
introduction of a long-range interaction, as has previously been
studied in the context of dipolar interactions \cite{cooper05}.

The paper is organised as follows.
In Sec.~\ref{sec:nonabelian} we introduce the non-Abelian gauge field. 
In Sec.~\ref{sec:interaction} we introduce the interaction Hamiltonian and the
corresponding Haldane pseudopotentials.
In Sec.~\ref{sec:vortexlattice} we present the vortex lattices.
In Sec.~\ref{sec:fqhs} we comment on fractional quantum Hall states in the system.
Sec.~\ref{sec:conclusions} contains our concluding remarks.
Some calculational details are relegated to Appendices.

\section{A non-Abelian gauge field}
\label{sec:nonabelian}

The Hamiltonian for a non-relativistic particle of mass $m$ and charge $q$ reads
\begin{equation}
H = \frac{1}{2m} (\vec{p} - q\vec{A})^2
\end{equation}
where $\vec{p}$ is the particle momentum and $\vec{A}$ is a vector potential which produces the magnetic field $\vec{B}=\vec{\nabla}\times\vec{A}$.
We are interested in the case of a non-Abelian vector potential which is written in the form
\begin{equation}  \label{eq:vectorpotential}
\vec{A} = A_x \vec{u}_x + A_y \vec{u}_y + A_z \vec{u}_z
\end{equation}
and the components $A_\mu$ are matrices acting on a set of states which, in the present context, correspond to a set of degenerate dressed states\cite{dalibard_gerbier_RMP2011}. In the simplest case there are two such internal states, which we label by $\sigma = \uparrow,\downarrow$, and $A_\mu$ are $2\times 2$ Hermitian matrices. The gauge group is then $U(2)=U(1) \times SU(2)$ and thus contains the standard $U(1)$ potential and a $SU(2)$ part.
The corresponding magnetic field is \cite{rubakov}
\begin{equation}  \label{eq:magneticfield}
\vec{B} = \vec{\nabla}\times \vec{A} - i \frac{q}{\hbar}\vec{A}\times\vec{A}.
\end{equation}
This relation shows that, in the case that the components of $\vec{A}$ are non-commuting, a nonzero magnetic field is obtained even from a uniform vector potential.

Let us consider a uniform magnetic field perpendicular to the plane $\vec{B}= B_z \vec{u}_z$, where $B_z$ is a $2\times 2$ hermitian matrix.
The magnetic field can be assumed to be diagonal by an appropriate choice of basis and we write $B_z = B ({\cal I} + 2 \beta^2 \sigma_z)$, where ${\cal I}$ is the identity matrix, $\sigma_z$ is the diagonal Pauli matrix, and $\beta$ is a parameter controlling the size of the non-Abelian part of the field.
The first Abelian term in such a magnetic field can be produced in the standard way by the rotation term in Eq.~\eqref{eq:magneticfield}. 
The second term of the magnetic field is produced by a vector potential whose components are non-commuting constant matrices, through the second term on the right-hand side of Eqn.~\eqref{eq:magneticfield}.
The complete non-Abelian vector potential may be chosen in the form
\begin{equation}
\vec{A} = B\, \left( \begin{array}{c} -y  {\cal I} \\ 0  \\ 0 \end{array} \right)
 + \beta'\, \left( \begin{array}{c} -\sigma_y  \\ \sigma_x  \\ 0 \end{array} \right)
\end{equation}
where $\sigma_x, \sigma_y$ are Pauli matrices. 
The constants in the magnetic field and the vector potential are related by $\beta = \beta'/(\ell_B B)$,
where we have introduced the magnetic length
$\ell_B \equiv \sqrt{\hbar/(q B)}$.
We have chosen the Landau gauge for the Abelian part, while
the second, non-Abelian term has been chosen to stand in analogy to the symmetric gauge.
Up to an overall gauge transformation 
this gauge potential is equivalent to the gauge potentials studied in 
Ref.~\cite{Burrello_Trombettoni_2010,estienne11} and for the symmetric case $|a|=|b|$ 
in Ref.~\cite{Palmer_Pachos_2011}. A method for implementing such a non-Abelian gauge field is described 
in Ref.~\cite{Burrello_Trombettoni_2011}.

\begin{figure}[t]
   \centering
   \includegraphics[width=2.8in]{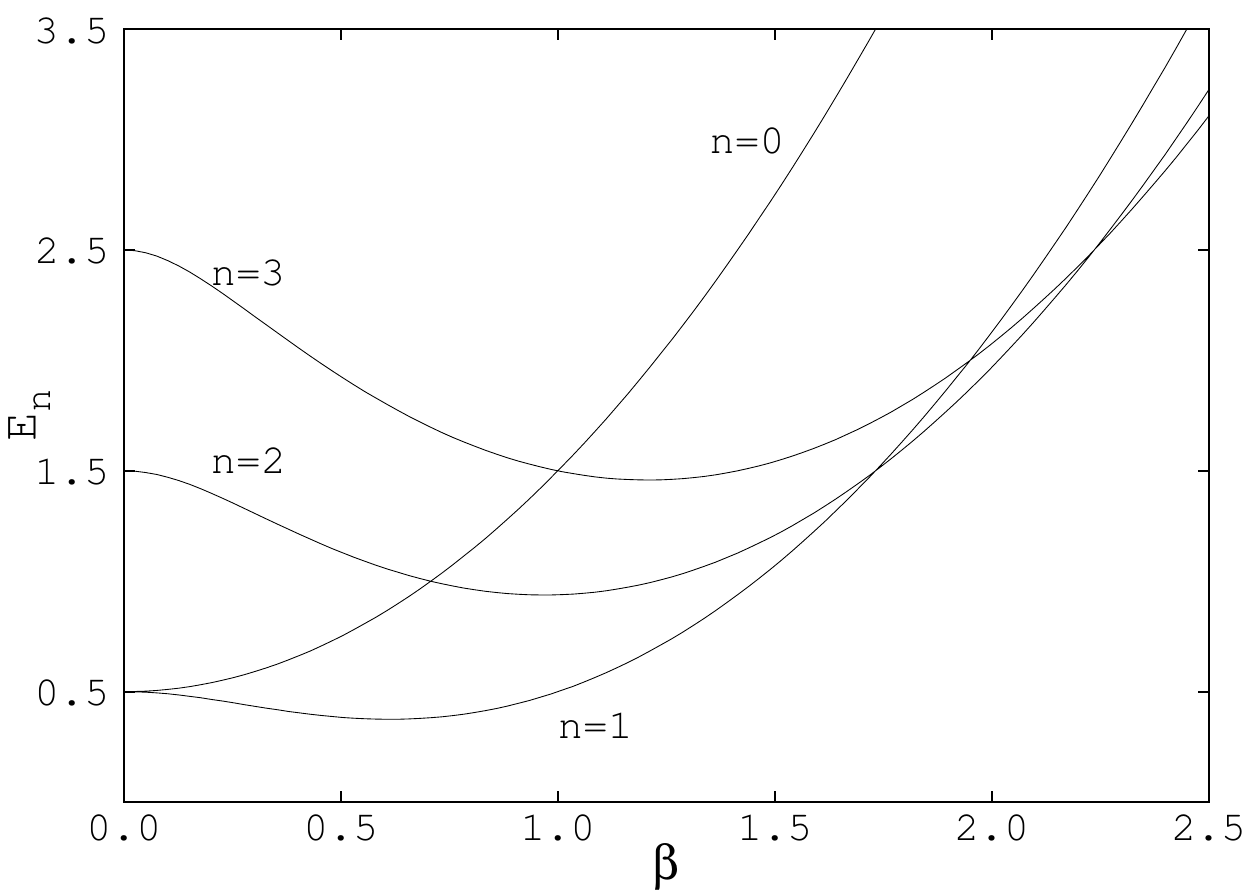}
   \caption{The energy levels \eqref{eq:energylevels} for $n=0,1,2,3$ as functions of $\beta$. We find that $E_1$ is the minimum energy for $0 < \beta < \sqrt{3}$ while $E_2$ is the minimum for $\sqrt{3} < \beta < \sqrt{5}$. }
   \label{fig: nonabelian_energy}
\end{figure}

We introduce the standard creation and annihilation operators $a^\dagger, a$ for the Landau level problem and the spin ladder operators $\sigma_\pm = (\sigma_x \pm \sigma_y)/2$, and then the Hamiltonian is written in the form
\begin{equation}  \label{eq:hamiltonian}
H = \hbar\omega_c\, \left[ a^\dagger a + \sqrt{2}\beta (a^\dagger \sigma_+ + a\,\sigma_-) + \frac{1}{2} + \beta^2 \right],
\end{equation}
where $\omega_c=qB/m$ is the cyclotron frequency.
This is equivalent to the Jaynes-Cummings Hamiltonian\cite{Shore_Knight_1993} and its spectrum is
\begin{align}  \label{eq:energylevels}
E_0 & = \hbar\omega_c \left( \frac{1}{2} + \beta^2 \right)  \nonumber \\
E_n^\pm & = \hbar\omega_c \left( n \pm \frac{\beta_n}{2} + \beta^2 \right),\quad \beta_n \equiv \sqrt{8\beta^2 n + 1},\; n \geq 1.
\end{align}
We are interested in the ground state as a function of the parameter $\beta$. Since $E_n^+ \geq E_n^-$ we will only discuss $E_n = E_n^-$. Fig.~\ref{fig: nonabelian_energy} shows the first few energy levels as a function of $\beta$.  One finds that $E_1$ is the lowest energy for $0 \leq \beta < \sqrt{3}$, while the ground state energy is $E_n$ for $\sqrt{2n-1} < \beta < \sqrt{2n+1}$ when $n>1$.

Let us denote the Landau level states by $\phi_{n,k}$ where $n$ is the Landau level index and $k$ is
an index for the degenerate states in each Landau level.
The two spin states will be denoted by the symbols $\up$ and $\down$ so the product states of the Landau level states with the spin states are denoted as $\phi_{n,k,\up},\; \phi_{n,k,\down}$.
The normalized eigenfunctions for the energies $E_n$ are
\begin{equation}  \label{eq:eigenfunction}
\psi_{n,k} = \left( \frac{\beta_n +1}{2\,\beta_n }\right)^{1/2} \phi_{n-1,k, \down}
 - \left( \frac{\beta_n -1}{2\,\beta_n} \right)^{1/2} \phi_{n,k, \up}.
\end{equation}
In  Appendix A, we give explicit expressions for the spatial dependence of the wavefunctions
$\phi_{n,k}$ in the periodic geometry used below for our numerical calculations.

\section{Interaction Hamiltonian}
\label{sec:interaction}

We assume that the bosons are interacting. We use a second-quantized description, and write $\psi_\sigma^\dagger(\vec{r})$ and $\psi_\sigma(\vec{r})$ for the creation and annihilation operators of a particle with spin $\sigma$ at position $\vec{r}$. The interaction Hamiltonian for two-body interactions has the form
\begin{equation}  \label{eq:inthamiltonian0}
H_I = \half \sum_{\sigma \sigma'} \int g_{\sigma\sigma'}(|\vec{r}-\vec{r}'|)\,
 \psi_{\sigma'}^\dagger(\vec{r}') \psi_\sigma^\dagger(\vec{r}) \psi_\sigma(\vec{r}) \psi_{\sigma'}(\vec{r}')\,
 d^2\vec{r}\, d^2\vec{r}'.
\end{equation}
We have assumed that the interaction potential depends on the distance between the particles and it may also depend on the spins. We have made the additional assumption that the spins of the bosons do not change due to scattering. We suppose a contact interaction potential of the form $g_{\sigma\sigma'}(|\vec{r}-\vec{r}'|) = g\, \delta(\vec{r}-\vec{r}')$ and obtain
\begin{equation}  \label{eq:inthamiltonian1}
H_I = \frac{g}{2}\,  \sum_{\sigma \sigma'} \int
 \psi_{\sigma'}^\dagger \psi_\sigma^\dagger \psi_\sigma \psi_{\sigma'}\,d^2\vec{r}.
\end{equation}
(It is straightforward to retain the spin-dependence to the interactions, which will introduce additional parameters to the model. We restrict attention to the spin-independent case to simplify presentation.)

We consider, first, the effect of interactions within the basis of single-particle eigenstates of definite angular momentum (\ref{eq:eigenfunction}). We assume that the interactions are small compared to the energy level spacings between states with different quantum numbers $n$, so we focus only on the scattering of pairs of particles with the same value of $n$, but between different angular momentum states: that is $(n,m_1),  \; (n,m_2)\to  (n,m_1'),  \; (n,m_2')$.
Owing to the rotational symmetry of the (contact) interactions, all two-body scattering processes (\ref{eq:inthamiltonian1}) preserve the total angular momentum of the pair of particles, $m_1'+m_2' =m_1+m_2$. As a result, the two-particle states with definite relative angular momentum, $m$, are eigenstates of the interaction Hamiltonian; their eigenvalues are the Haldane pseudopotentials\cite{haldanesphere,jain}, $V^{(n)}_{m}$, which encode the entire properties of the interactions within the energy band labelled by $n$. For the states (\ref{eq:eigenfunction}) 
these take the form
\begin{align}  \label{eq:pseudo_pot}
V^{(n)}_m = \frac{g}{4\pi} \int_0^\infty & \left[\frac{\beta_n+1}{2\beta_n} L_{n-1}(q^2/2)
+\frac{\beta_n-1}{2\beta_n} L_{n}(q^2/2)\right]^2  \nonumber \\
  & \left[L_m(q^2)\right]^2 e^{-q^2} \;d q^2.
\end{align}
We use here and in the following as a unit of length the magnetic length $\ell_B$.

For bosons, symmetry of the two-particle wave functions means that only even $m=0,2,4,\ldots$ can contribute.
For the usual case of rapidly rotating bosons in the lowest Landau level, $n=0$, interacting via contact interactions \cite{cooper08}, the only non-zero pseudopotential is $V_0$. This represents the situation in which interactions have the shortest possible range. Explicitly introducing a long-range interaction potential, such as the dipolar potential with $g(r) \propto 1/r^3$ at large distances, leads to non-zero values of pseudopotentials with $m>0$\cite{cooper05}.  We find, from an analysis of (\ref{eq:pseudo_pot}), that when the non-Abelian gauge field is present, $\beta\neq 0$, a set of pseudopotentials with $m > 0$ are also non-zero even for contact interactions $g(|\vec{r}|) \propto \delta(\vec{r})$. We therefore argue that {\it the effect of the non-Abelian gauge field on the properties of the lowest energy state is equivalent to the introduction of effective interactions with non-zero range.}

In the case of dipolar interactions, these changes to the effective interaction were found to lead to changes in the lowest energy vortex lattice phase \cite{cooper05}. In the following we investigate the effects on the vortex lattices due to the changes in effective interaction caused by the non-Abelian gauge field.
We first give results of numerical simulations and we then make the connection to the calculated values of the pseudopotentials  $V^{(n)}_m$.

\section{Vortex lattice phases}
\label{sec:vortexlattice}

To study vortex lattices of an infinite system, it is convenient to
work in a rectangular geometry with periodic boundary conditions in
the two spatial co-ordinates, that is with the topology of a torus.
We consider a rectangular cell of size $L_x$ and $L_y$ in the $x$ and $y$ directions
respectively. This spatial periodicity  imposes the condition
\begin{equation}  \label{eq:fluxquanta}
L_x L_y = 2\pi\ell_B^2\, \Nv,
\end{equation}
where $\Nv$ is a positive integer equal to the number of states in the
cell for each Landau level. This can be interpreted as the number of
vortices within the cell.  
Depending on the symmetry of the vortex lattice phase (described
within mean field theory below),
the aspect ratio of the cell $L_y/L_x$ must be chosen
to match the natural periods of the vortex lattice.

 For small enough values of the interaction strength $g \bar{\rho} \ll \hbar \omega_c$, where $\bar{\rho} \equiv N/L_xL_y$ is the mean particle density, we can assume that the interaction potential does not mix excited states of the non-interacting Hamiltonian. Therefore the ground state of the interacting system is found by minimizing the interaction energy within the space of the wave functions \eqref{eq:eigenfunction} for a specific $n$.
 
We make the ansatz
\begin{equation}  \label{eq:ansatz}
\Psi_n = \sum_k c_k \psi_{n,k}
\end{equation}
where $c_k$ are complex coefficients, and $\psi_{n,k}$ are the single particle wavefunctions (\ref{eq:eigenfunction}) for the periodic geometry (see Appendix A). The choice for the Landau level index $n$ depends on the value of $\beta$. 
We minimize the energy
\begin{equation}  \label{eq:interactionenergy}
E_I \equiv \frac{g}{2}\;  \sum_{k_1,k_2,k_3,k_4} V_{k_1,k_2,k_3,k_4}\; c_{k_1}^* c_{k_2}^* c_{k_3} c_{k_4}
\end{equation}
(the form of the interaction potentials $V_{k_1,k_2,k_3,k_4}$ is given in Appendix \ref{app:interactionpotentials})
in the subspace of wave functions \eqref{eq:ansatz}
for a given number of particles
\begin{equation}  \label{eq:numberparticles}
N = \sum_k c_k^* c_k.
\end{equation}
The mean density is then $\bar{\rho} = \frac{N}{N_V}\frac{1}{2\pi \ell_B^2}$, and the condition that mean-field theory provides an accurate description is that $N/N_{\rm V} \gg 1$ \cite{cooper08}.

We find minima of the energy \eqref{eq:interactionenergy} numerically using the Fletcher-Reeves-Polak-Ribiere method \cite{numericalrecipes} which uses derivatives of the function to be minimized. 
The number of basis wave functions in the ansatz \eqref{eq:ansatz} is $\Nv$ and we used values up to $\Nv = 64$ for which the method rapidly converges to a minimum.
Setting $\Nv$ to a certain value we fix the number of Landau level states 
which are introduced in the area $[L_x,0]\times [0,L_y]$, and the cell area is fixed through Eq.~\eqref{eq:fluxquanta}.
As a second parameter we choose the aspect ratio $a_r = L_y/L_x$ of the rectangular cell.
We have explored the minima of the interaction energy in the space of the two parameters $\Nv,\,a_r$. 
For every state under consideration we evaluate the quantity
\begin{equation}  \label{eq:bfactor}
b \equiv \frac{E_I}{N^2}\, (4\pi \Nv),
\end{equation}
where $E_I$ is the interaction energy and $N$ is the number of particles in a cell with $\Nv$ vortices,
from which the chemical potential is $\mu = \frac{dE}{dN} = b \bar{\rho}$.

For $\beta=0$ the ground state wave functions are the lowest Landau level states $\phi_{0,k,\down}$, and we can verify that a triangular vortex lattice (identical to Fig.~1a in Ref.~\cite{cooper06}) is the lowest energy state with $b=1.596$, as expected by the equivalent problem in the Ginsburg-Landau model for type-II superconductors \cite{abrikosov57,kleiner_roth_autler_PR64}.

\begin{figure}[t]
   \centering
  \includegraphics[width=3.4in]{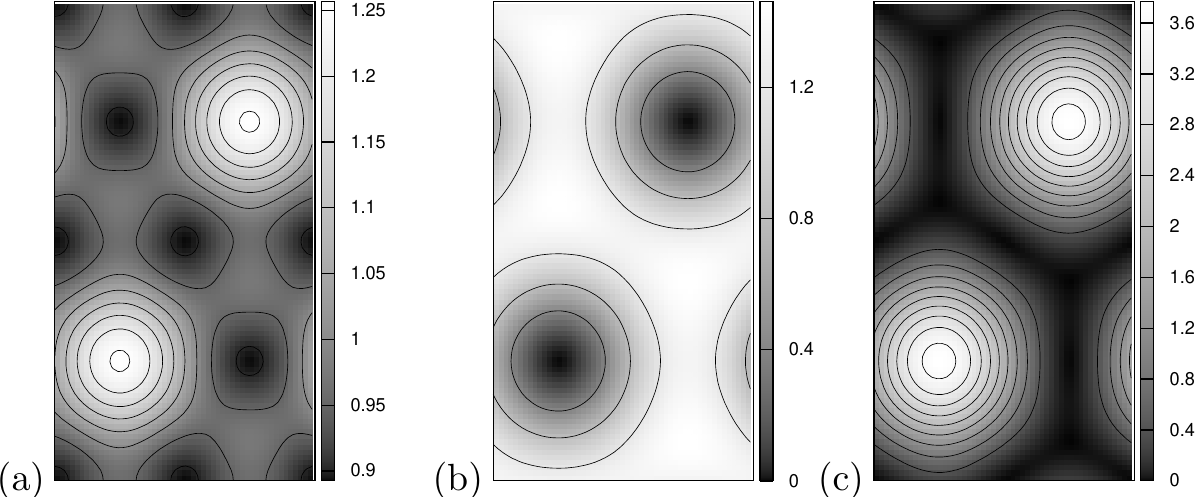}
   \caption{Contour plots for the particle density of a triangular vortex lattice for $\beta=1.0$
   (aspect ratio $a_r=\sqrt{3}$).
   We present one cell which contains $\Nv=2$  vortices.
   (a) The total particle density (in units of the mean particle density).
       The particle density is nowhere zero, since the 
 single particle wavefunctions involve contributions from more than one Landau level, as shown in Eqs.~\eqref{eq:eigenfunction}, \eqref{eq:ansatz}.
   (b) The density for particles in the lowest Landau level (in units of the mean particle density in the lowest Landau level).
   (c) The density for particles in the first Landau level (in units of the mean particle density in the first Landau level).
   }
   \label{fig:beta100}
\end{figure}

For $0 < \beta < \sqrt{3}$ the ground state wave functions are obtained by setting $n=1$ in Eq.~\eqref{eq:eigenfunction}, and they are superpositions of the lowest and first Landau level states.
Minimizing \eqref{eq:interactionenergy} we find a triangular lattice as the lowest energy wave function for the range $0 \leq \beta \leq 1.13$. 
Note, however, that the groundstate wave function depends on $\beta$ explicitly through the coefficients of $\phi_{0,k,\down}$ and $\phi_{1,k,\up}$ in Eq.~\eqref{eq:eigenfunction}.
In Fig.~\ref{fig:beta100} we present plots of the particle density for $\beta=1.0$.
We plot the total density as well as the densities for the two Landau levels of the wavefunction.
As $\beta$ increases the contribution of $\phi_{1,k,\up}$ in the wave function increases as indicated in Eq.~\eqref{eq:eigenfunction}.
The contribution to the particle density from the first Landau level wave function $\phi_{1,k,\up}$ has maximum values where the vortex centers for $\phi_{0,k,\up}$ are located.
These remarks suffice in order to follow the change of the particle density in the lattice as $\beta$ increases.
Note that we have density minima, but no zeros of the density, in Fig.~\ref{fig:beta100}a.

\begin{figure}[t]
   \centering
     \includegraphics[width=3.4in]{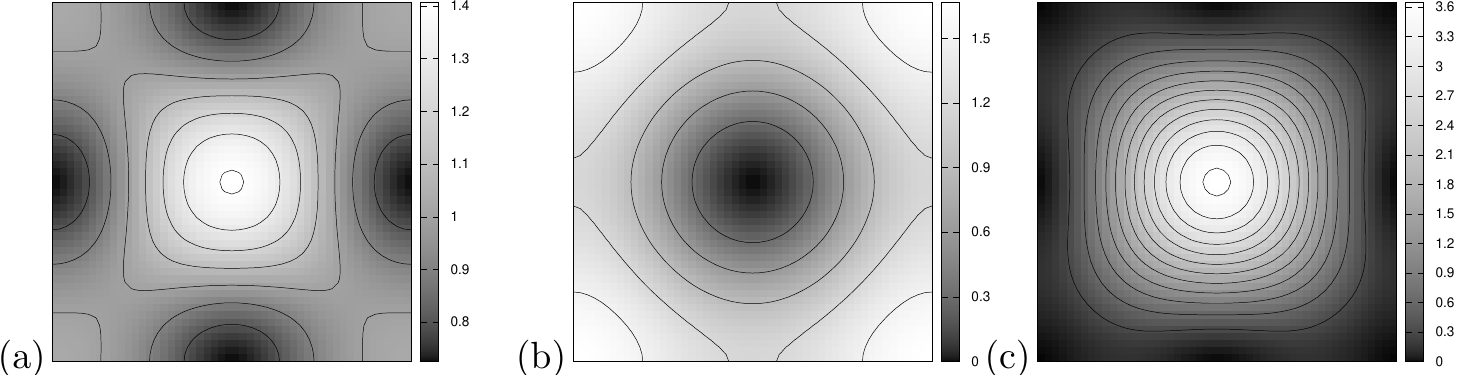}
   \caption{Contour plots for the particle density of a square vortex lattice for $\beta=1.5$
   (aspect ratio $a_r=1$). We present one cell which contains $\Nv=1$  vortex.
   (a) The total particle density.
       The particle density is nowhere zero, since the 
 single particle wavefunctions involve contributions from more than one Landau level.
   (b) The density for particles in the lowest Landau level.
   (c) The density for particles in the first Landau level.
}
   \label{fig:beta150}
\end{figure}

For $1.13 < \beta < \sqrt{3}$ our numerical method converges to a square lattice which has lower energy than any other state we investigated. 
 Fig.~\ref{fig:beta150} shows the result of the energy minimization in a single unit cell for a square lattice for $\beta=1.5$.

A transition from a triangular to square and other vortex lattices has been investigated for the case of long-range dipolar interactions in the LLL \cite{cooper05}. In order to quantify the contribution of non-local interactions we calculate the values of pseudopotentials $V^{(n)}_m$.
It is convenient to define as a control parameter the ratio of the two first nonzero pseudopotentials at a certain Landau level \cite{cooper06}
\begin{equation}  \label{eq:alpha}
\alpha \equiv \frac{V_2}{V_0}.
\end{equation}
We have simplified the notation setting $V_m=V_m^{(1)}$.
If $V_m=0$ for $m>2$ a transition from triangular to square lattice occurs at $\alpha = \alpha_1 \equiv 0.0865$ \cite{cooper06}.
For $n=1$ in the present model we find that $V_m = 0$ for $m > 2$, and the value $\alpha= \alpha_1$ is obtained for $\beta=1.13$.
Therefore, the transition to a square lattice for $\beta > 1.13$ reported here is in perfect agreement with the results of Ref.~\cite{cooper06}.
The parameter $\alpha$ increases with $\beta$, and for $\beta=\sqrt{3}$ we have $\alpha=1.118$.
In further agreement of the present results to the results of Ref.~\cite{cooper06}, no transition to a vortex phase different than the square lattice is expected for these parameter values.

\begin{figure}[t]
   \centering
     \includegraphics[width=3.4in]{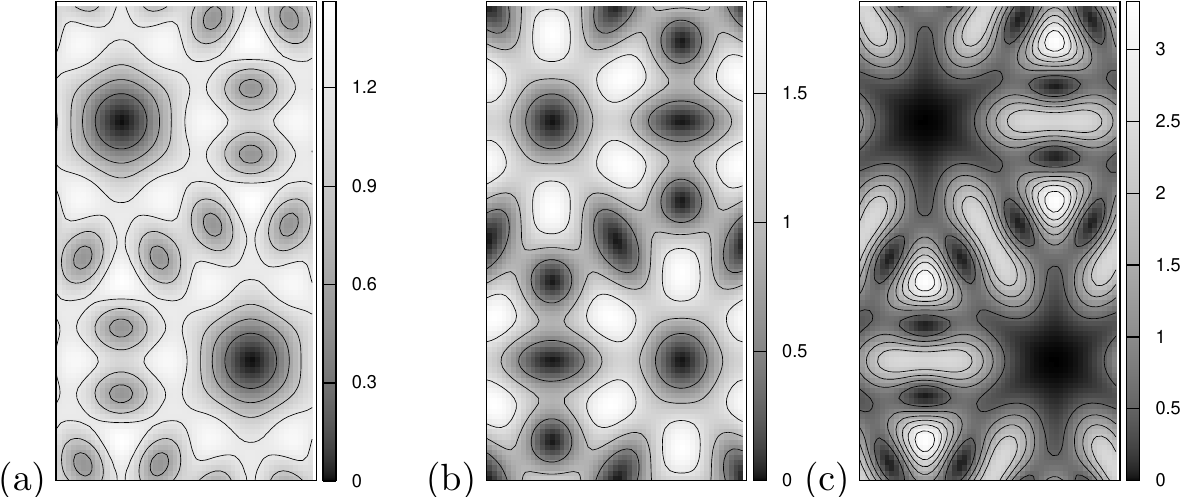}
      \caption{Contour plots for the particle density of a $q=4$ bubble crystal lattice for $\beta=2.0$
   (aspect ratio $a_r=\sqrt{3}$). We present a cell which contains $\Nv=8$  vortices.
   (a) The total particle density.
       The particle density is zero on the sites of a triangular lattice.
   (b) The density for particles in the first Landau level.
   (c) The density for particles in the second Landau level.
   }
   \label{fig:beta200}
\end{figure}

When $\beta$ increases to values $\sqrt{3} < \beta < \sqrt{5}$ the ground state of the non-interacting problem is  a superposition of the first and second Landau levels obtained from wave function \eqref{eq:eigenfunction} with $n=2$. 
 Fig.~\ref{fig:beta200} shows the particle density of the ground state for $\beta=2.0$.
The picture is similar for the whole range of values of $\beta$. The particle density 
has vortex minima arranged in a triangular lattice and these are surrounded by density dips.
The clusters of vortex and density dips may be called {\it bubbles} and we thus have a bubble crystal phase in Fig.~\ref{fig:beta200}.

Following Refs.~\cite{cooper05, cooper06} we may characterize the vortex lattice in Fig.~\ref{fig:beta200} as a $q=4$ bubble as each unit cell contains 4 vortices.
The particular bubble phase of Fig.~\ref{fig:beta200} is similar to that in Fig.~1e in Ref.~\cite{cooper06}.
The Haldane pseudopotentials for $\beta = \sqrt{3}$ are $\alpha = V_2/V_0=0.65$ and $V_4/V_0=0.16$,
where we use the simplified notation $V_m = V_m^{(2)}$.
The vortex lattice for this value of $\alpha$ gives a $q=4$ bubble state in the case of dipolar interactions \cite{cooper05} as well as in the model of Ref.~\cite{cooper06}.
The present results are thus in agreement with the preceding reports showing that the non-Abelian gauge potential has an effect analogous to long-range interactions on the vortex lattice phases.

Note that we find no evidence of the appearance of
  ``stripe crystal'' phases, which appear for $0.12 \leq
  \alpha \leq 0.70$ for the model of Ref.~\cite{cooper06}. For
  the present model we have shown that $0<\alpha < 0.12$ for $0<
  \beta < \sqrt{3}$ and $\alpha > 0.70$ for $\beta > \sqrt{3}$ as a
  result of the transition for the groundstate in the noninteracting model
  at $\beta=\sqrt{3}$. A direct transition from square lattice to
  bubble crystal at $\beta = \sqrt{3}$ is therefore consistent with
  the results of Ref.~\cite{cooper06}.

\section{Consequences for Fractional Quantum Hall States}
\label{sec:fqhs}

The above mean-field studies are valid in the regime of high filling
factor $\nu \equiv N/N_{\rm V}\gg 1$ \cite{cooper08}.  For small
values of the filling factor, the vortex lattice phases can be
replaced by strongly correlated phases, which are bosonic analogues of
fractional quantum Hall (FQH) states. For atoms in a uniform Abelian
gauge field at $\nu=1/2$ and interacting with contact interactions,
the groundstate at $\nu=1/2$ is the Laughlin
state\cite{wilkin_gunn_smith98}, and those at $\nu=1,3/2,2$ are well
described by the Moore-Read and Read-Rezayi states\cite{cooper_wilkin_gunn_PRL2001}.
Variations in the (ratios of the) Haldane pseudopotentials can lead to
changes in the nature of the groundstate.

Our result, Eqn (\ref{eq:pseudo_pot}), provides the values of the
pseudopotentials for the non-Abelian gauge potential studied here.  By
combining these results with existing numerical studies of the effects
of variations of the pseudopotentials on strongly correlated phases\cite{cooper05,cooper06,rrc,regnaultjolicoeur07}
we can deduce consequences of the non-Abelian gauge field for the FQH
states.

For filling factor $\nu=1/2$ the effect of changing the Haldane
pseudopotentials was studied in
Refs.~\cite{cooper05,cooper06}. For a model with only the lowest
two pseudopotentials, with ratio $\alpha = V_2/V_0$, it was found that
the groundstate is described by the bosonic Laughlin state for $\alpha
\lesssim 0.4$ \cite{cooper06}. For $\alpha \gtrsim 0.4$ the
groundstate is a compressible crystalline phase related to the bubble
crystal phase of mean-field theory described above. For larger filling
factors, $\nu > 1/2$ the crystalline phase becomes increasingly
stable.  Thus, for $\alpha \gtrsim 0.4$ the groundstates at all
filling factors $\nu\geq 1/2$ are compressible crystalline states.
This has been confirmed in numerical studies for filling factors
$\nu=1,3/2, 2$\cite{rrc,cooper_rezayi_PRA2007,regnaultjolicoeur07}.

In the present model, with a non-Abelian gauge field, we have shown
that for $0\leq \beta \leq \sqrt{3}$ the lowest energy single particle
state has $n=1$ and only the pseudopotentials $V_0$ and $V_2$ are
nonzero. The ratio $\alpha = V_2/V_0$ is in the range $0\leq \alpha
\leq 0.118$. Thus, over this range, one expects the groundstate at
$\nu=1/2$ to be an incompressible liquid that is well described by the
Laughlin state\cite{cooper06}. Similarly, over this range, $\alpha$ is
sufficiently small that the groundstates at $\nu=1,3/2$ and  $2$ are well
described by the Moore-Read state, the $k=3$ Read-Rezayi state, and the $k=4$
Read-Rezayi state respectively\cite{cooper_wilkin_gunn_PRL2001}. Indeed, a small non-zero
value of $\alpha$ has been shown to tune the system into a regime
where these very interesting non-Abelian quantum Hall phases describe
the groundstate accurately\cite{rrc,cooper_rezayi_PRA2007,regnaultjolicoeur07}.

For $\sqrt{3} < \beta < \sqrt{5}$, the lowest energy state has
$n=2$. The non-zero pseudopotentials are $V_0, V_2$ and $V_4$.  Over
this range of $\beta$, the ratio $\alpha = V_2/V_0 \simeq 0.65$, while
$V_4/V_0 \lesssim 0.2$.  Neglecting the effect of this small value
of $V_4$, we can again make use of the results for the pure
$V_0$-$V_2$ model.  (Neglecting $V_4$ was accurate for
interpreting the mean-field groundstate described above.) Now, the
ratio $\alpha = 0.65$ is so large ($> 0.4$) that the groundstate at
all filling factors $\nu\geq 1/2$ is expected to be a compressible
crystalline
state\cite{cooper06,rrc,cooper_rezayi_PRA2007,regnaultjolicoeur07}.

Thus we can conclude that, for the filling factors
$\nu=1/2,1,3/2,2$, there is a transition from incompressible quantum
liquid states (Laughlin and Read-Rezayi states) to a compressible
crystalline state as the non-Abelian field is increased through $\beta
= \sqrt{3}$.  This conclusion is in agreement with independent exact
diagonalization calculations reported in Ref.\cite{Palmer_Pachos_2011} in those
parameter regimes for which there is overlap.

\section{Conclusions}
\label{sec:conclusions}

We have studied vortex phases in a model for charged particles in a non-Abelian gauge potential pertaining to a $U(2)$ symmetry. Applying Gross-Pitaevskii mean field theory we have considered the effect of contact interactions between particles. These lead to the formation of triangular, square, and bubble crystal lattices for increasing values of the parameter for the non-Abelian term in the gauge potential.
We calculated  the Haldane pseudopotentials for the energy states of the non-Abelian model
and find that they indicate effective interactions with non-zero range in the system.
We find a general agreement with results on the effect of long-range interactions \cite{cooper05,cooper06}.
We conclude that the effect of the non-Abelian gauge field on the properties of the lowest energy state is equivalent to the introduction of effective interactions with non-zero range.

We suppose throughout that the interactions are small compared to Landau level spacing. However, at values for the parameter $\beta \sim \sqrt{2n+1}$,  where the successive energy levels for non-Abelian model cross, our approximation is not valid, since Landau level mixing \cite{komineas_cooper_PRA2007} is then expected even for small interactions.

\section*{Acknowledgements}

S.K. is grateful to the TCM Group of the Cavendish Laboratory for hospitality. This work was partially supported by the FP7-REGPOT-2009-1 project ``Archimedes Center for Modeling, Analysis and Computation'' and by EPSRC Grant EP/F032773/1.

\appendix

\section{Spatially periodic wave functions}
\label{app:normalization}

The single-particle energy eigenstates of the non-Abelian gauge
  field are given by Eqn. (\ref{eq:eigenfunction}),
  in which $\phi_{n,k}$ are the normalized Landau level wavefunctions for uniform
  Abelian magnetic field. For the periodic system we study, with a
  rectangular cell of size $L_x\times L_y$, these functions
  must have spatial periods of $L_x$ and $L_y$ in the $x$ and $y$ directions.

The lowest Landau level wave functions with
  these periodicities are \cite{yoshioka83}
\begin{align}  \label{eq:lll}
& \phi_{0,k}= \frac{1}{(L_x \pi^{1/2})^{1/2}}\; \sum_{p=-\infty}^{\infty} e^{i (Y_k + p L_y)x}\, e^{-\half (Y_k + p L_y-y)^2},  \\
& Y_k \equiv \frac{2\pi}{L_x} k, \nonumber
\end{align}
where $k$ takes the integer values $k=0,\ldots,\Nv-1$, with $\Nv$ given
by Eqn. (\ref{eq:fluxquanta}).
Similarly, the spatially periodic wave functions in the first and second Landau levels are
\begin{align}  \label{eq:1ll}
\phi_{1, k} =
 \frac{1}{(L_x\, 2\pi^{1/2})^{1/2}}\; \; \sum_{p=-\infty}^{\infty} &   2(Y_k + p L_y-y)\, e^{i (Y_k + p L_y)x}  \nonumber \\
      &    e^{-\half (Y_k + p L_y-y)^2},
\end{align}
\begin{align}  \label{eq:2ll}
\phi_{2, k} =
 \frac{1}{(L_x\, 8\pi^{1/2})^{1/2}}\; \; \sum_{p=-\infty}^{\infty} &  [4(Y_k + p L_y-y)^2-2]  \nonumber  \\
              & e^{i (Y_k + p L_y)x}\, e^{-\half (Y_k + p L_y-y)^2}.
\end{align}

These wavefunctions are orthonormal within the periodic cell. Consider, for example, the wave functions \eqref{eq:lll}.
We have
\begin{align}  \label{eq:orthonormality1}
\int_0^{L_x} \int_0^{L_y} & \phi_{0,k_1}^* \phi_{0,k_2}\, dx dy =  \\ 
 & \sum_{p_1, p_2} \frac{\delta_{k_1+p_1 \Nv,k_2+p_2 \Nv}}{\pi^{1/2}}\, \int_0^{L_y} e^{-(Y_k + p L_y-y)^2}\,dy, \nonumber
\end{align}
where all summations in the $p$ symbols (here and in the following) extend from $-\infty$ to $\infty$.
The $\delta$ symbol in the latter equation gives zero for every $p_1 \neq 0 \neq p_2$ since the $k's$ take values in the range $0 \leq k_1, k_2 < \Nv$.
Therefore it is equal to the product $\delta_{k_1,k_2}\,\delta_{p_1,p_2}$.
Using the result
\begin{equation}  \label{eq:periodic_integral}
\sum_p \int_0^{L_y} dy\; e^{-(Y_m + p L_y-y)^2} = \int_{-\infty}^{\infty} dy\, e^{- \left(y - Y_m \right)^2} = \sqrt{\pi},
\end{equation}
which we substitute in Eq.~\eqref{eq:orthonormality1}, we finally find that $\phi_{0,k}$ are orthonormal. A similar procedure for the first ($n=1$) and second ($n=2$) Landau level wave functions proves the orthonormality condition
\begin{equation}  \label{eq:orthonormality}
\int_0^{L_x} \int_0^{L_y} \phi_{n,k_1}^* \phi_{n,k_2}\, dy dx = 
\delta_{k_1, k_2}.
\end{equation}

\section{Interaction potentials}
\label{app:interactionpotentials}

The interaction potentials entering in the calculation of the interaction energy \eqref{eq:interactionenergy} are of the form
\begin{align}  \label{eq:potential}
V_{k_1,k_2,k_3,k_4}
 = & \left( \frac{\beta_n +1}{2\beta_n} \right)^2
  \Big[ V_{k_1,k_2,k_3,k_4}^{\down\;\;\;\down\;\;\;\down\;\;\;\down}  \nonumber \\
& + \frac{\beta_n -1}{\beta_n +1}\, (V_{k_1,k_2,k_3,k_4}^{\up\;\;\;\down\;\;\;\down\;\;\;\up}
+ V_{k_1,k_2,k_3,k_4}^{\down\;\;\;\up\;\;\;\up\;\;\;\down} ) \nonumber \\
& + \left( \frac{\beta_n -1}{\beta_n +1}\right)^2 V_{k_1,k_2,k_3,k_4}^{\up\;\;\;\up\;\;\;\up\;\;\;\up}
\Big],
\end{align}
where
\begin{align*}
V_{k_1,k_2,k_3,k_4}^{\down\;\;\;\down\;\;\;\down\;\;\;\down} & \equiv 
 \int \phi_{n-1,k_1}^* \phi_{n-1,k_2}^* \phi_{n-1,k_3} \phi_{n-1,k_4}\,d^2\vec{r},  \\
V_{k_1,k_2,k_3,k_4}^{\up\;\;\;\up\;\;\;\up\;\;\;\up} & \equiv \int \phi_{n,k_1}^* \phi_{n,k_2}^* \phi_{n,k_3} \phi_{n,k_4}\,d^2\vec{r},  \\
V_{k_1,k_2,k_3,k_4}^{\up\;\;\;\down\;\;\;\down\;\;\;\up} & \equiv \int \phi_{n,k_1}^* \phi_{n-1,k_2}^* \phi_{n-1,k_3} \phi_{n,k_4}\,d^2\vec{r},  \\
V_{k_1,k_2,k_3,k_4}^{\down\;\;\;\up\;\;\;\up\;\;\;\down} & \equiv \int \phi_{n-1,k_1}^* \phi_{n,k_2}^* \phi_{n,k_3} \phi_{n-1,k_4}\,d^2\vec{r},
\end{align*}
and we have explicitly taken into account that the outgoing particles ($k_1, k_2$) have the same spin as the incoming ones ($k_3, k_4$). We present results for $n=1$ and $n=2$ in this paper.

\end{document}